\documentclass[conference]{IEEEtran}

\usepackage{cite}
\usepackage{amsmath,amssymb,amsfonts,amsthm}
\usepackage{algorithm} 
\usepackage{algorithmicx}
\usepackage{algpseudocode}

\usepackage{graphics,
	psfrag,
	epsfig,}

\usepackage{graphicx}
\usepackage{epstopdf}

\usepackage{tikz-cd}
\usepackage{epstopdf}

\newcommand{\sref}[1]{Section~\ref{#1}}

\newcommand{\cref}[1]{Constraint~\ref{#1}}

\usepackage[labelformat=simple]{subcaption}

\usepackage{graphicx}
\usepackage{textcomp}
\usepackage{xcolor}
\usepackage{tikz}
\usetikzlibrary{shapes,arrows}
\usepackage{booktabs}

\newcommand{\ignore}[1]{}
\def\BibTeX{{\rm B\kern-.05em{\sc i\kern-.025em b}\kern-.08em
    T\kern-.1667em\lower.7ex\hbox{E}\kern-.125emX}}
\begin{document}

\title{
Improving Connectivity of RIS-Assisted UAV Networks using RIS Partitioning and Deployment 

}

\author{\IEEEauthorblockN{Mohammed Saif and Shahrokh Valaee}
\IEEEauthorblockA{Department of Electrical and Computer Engineering, University of Toronto, Toronto, Canada\\ 
Email: mohammed.saif@utoronto.ca, valaee@ece.utoronto.ca
\vspace{-0.35cm}
\thanks{This work was supported in part by funding from the Innovation for Defence Excellence and Security (IDEaS) program from the Department of National Defence (DND) of Canada.}
}}

\IEEEoverridecommandlockouts
\maketitle

\IEEEpubidadjcol

\begin{abstract}
Reconfigurable intelligent surface (RIS) is pivotal for beyond 5G networks in regards to the surge demand for reliable communication in  unmanned aerial vehicle (UAV) networks. This paper presents an innovative  approach to maximize connectivity of UAV networks  using RIS deployment and virtual partitioning, wherein an RIS is deployed to assist in the communications between an user-equipment (UE) and blocked UAVs. Closed-form (CF) expressions for signal-to-noise ratio (SNR) of the two-UAV setup are derived and validated. Then, an optimization problem is formulated to maximize network connectivity by optimizing the 3D
deployment of the RIS and its partitioning subject to predefined quality-of-service (QoS) constraints.
To tackle this problem, we propose a method of virtually partitioning the RIS given a fixed 3D location, such that the partition phase shifts are configured to create cascaded channels between the UE and the blocked two UAVs. Then, simulated-annealing (SA) method is used to find the  3D location of the RIS. Simulation results
 demonstrate that the proposed joint
 RIS deployment and partitioning framework can significantly improve network connectivity compared to benchmarks, including RIS-free and RIS with a single narrow-beam link.

\end{abstract}

\begin{IEEEkeywords}
Network  connectivity, RIS-assisted UAV communications, RIS partitioning and deployment.
\end{IEEEkeywords}

\section{Introduction}
The proliferation of wireless devices, e.g., user-equipments (UEs) and unmanned aerial vehicles (UAVs), all requiring ultra-reliable connectivity, is on a rapid rise. Although UAVs have distinctive features, such as maintaining direct connections with UEs, rapid deployment, and adjustable mobility, the inherent blockage  of wireless channels between UEs and UAVs remains a persistent challenge, particularly in dense urban scenarios \cite{8292633, UAV_economy}. Furthermore, the high volume data of the unconnected UEs may exacerbate this issue, as critical data may not be reliably and quickly delivered to the UAVs and to the fusion center for processing. A possible solution to this connectivity problem is implementing relays \cite{9593204, 4657335, 4786516} or more UAVs,  but it is not always feasible to deploy more UAVs or relays due to site constraints, cost, and power consumption. In addition, UAVs are prone to failure due to limited energy. 

The emergence of reconfigurable intelligent surface (RIS) has introduced a revolutionary technology for  improving localization \cite{M}, energy efficiency \cite{Javad_globecom2023}, and coverage \cite{RIS_Mohanad, 9293155}. With its low-cost
passive reflecting elements, an RIS can be managed by a dedicated controller to adjust the electromagnetic
properties of incident signals,
thus influencing the signal’s strength and direction. Unlike other solutions, an RIS is an attractive technology due to desired properties such as 1) easy installation, 2) low cost, 3) passive elements, and 4) low energy consumption.

The integration of RIS and UAV communications has been considered in the literature for different purposes, such as maximizing energy efficiency \cite{Javad_globecom2023}, improving connectivity of UAV networks \cite{saifglobecom}, improving sensing and localization \cite{Ali}, enhancing secrecy rate of aerial-RIS networks \cite{PLS} and smart cities \cite{Mu}. In this paper, we are interested in maximizing connectivity of UAV networks via a joint optimization of RIS deployment and partitioning, which has not been considered in the literature. Specifically, in \cite{saifglobecom}, the authors utilize the RIS to create a reflected link to the desired UAV to improve connectivity of UAV networks using convex relaxation. The recent work \cite{PLS} enables the RIS to amplify the intended legitimate UE’s signal while attenuating the illegitimate signal to enhance the network secrecy via RIS partitioning. Additionally, \cite{aydin} uses the RIS to boost the signals' strength for resilient wireless networks.  None of these recent works have considered maximizing connectivity of UAV networks via RIS deployment and partitioning. 

Therefore, this work
presents the initial results for enhancing connectivity of UAV networks by leveraging the additional paths introduced by the cascaded channels of the RIS virtual representation. We present a closed-form analytical solution and a 3D RIS deployment approach for connectivity of UAV networks. Our proposed virtual RIS partitioning and deployment scheme enables the RIS to reflect the intended UE's signal to the blocked UAVs based on their reliability and quality-of-service (QoS) constraints. The obtained results illustrate that the proposed RIS partitioning model achieves superior connectivity compared to the conventional benchmarks, including the cases of RIS-free network and when the RIS aligns all its elements to one blocked UAV, creating a single narrow-beam link.

\section{System Model}\label{S}

\subsection{Network Model}
\begin{figure}[t!]  

\begin{center}
\includegraphics[width=0.75\linewidth, draft=false]{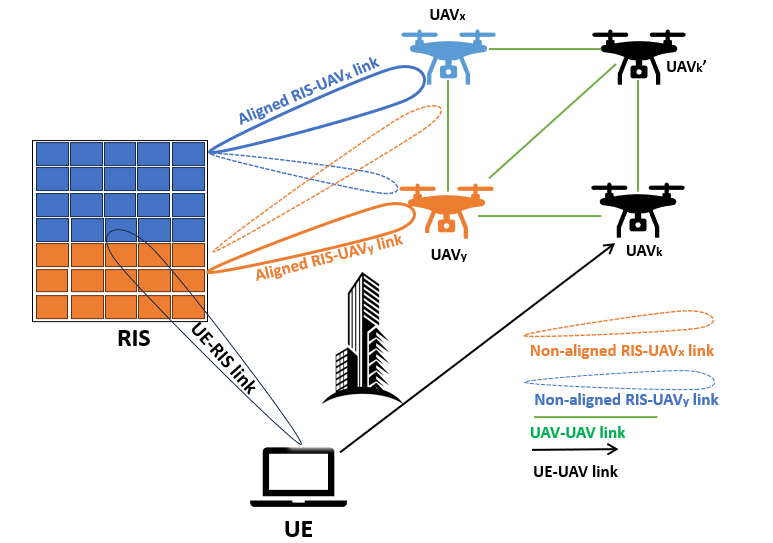}
\caption{Uplink RIS-assisted UAV model with one partitioned RIS, where RIS elements are virtually partitioned and coherently aligned (configured)
with UAV$_x$ and UAV$_y$.}
   \label{fig1}
			\end{center}
 
		\end{figure}

This paper presents an innovative approach to maximizing connectivity of UAV networks  using RIS deployment and partitioning,  as shown in Fig. \ref{fig1}. In the illustrated uplink RIS-assisted UAV scenario, one user-equipment (UE) intends to transmit data to the UAVs, denoted by the set $\mathcal K=\{1, 2, \ldots, K\}$, and RIS can aid in establishing reliable communications and support the direct link of the UE. Assuming a dense urban scenario, where direct links between the UE and some UAVs are blocked, and communication can only occur through the RIS, RIS can aid in establishing reliable communication with the blocked UAVs. To enable the RIS to coherently reflect the signal of the UE to multiple UAVs, we propose a virtual RIS partitioning approach, wherein a dedicated portion of RIS elements are configured to passively beamform the UE's signal to one UAV. We assume that the UE and the UAVs are equipped with a single antenna, while the RIS has $N$ reflecting elements that are indexed row-by-row by $n=1, \ldots, N$. The UE and UAVs transmit at identical powers, denoted by $p$ and $P$, respectively.

In the depicted scenario of Fig. \ref{fig1}, we consider the setup of two UAVs $x$ and $y$, represented by UAV$_{j}$, $j\in \{x,y\}$, which are blocked and can not communicate directly with the UE and must communicate through the RIS, where UAV$_x$ is more reliable than UAV$_y$. Low reliable UAV is the most critical one that causes severe connectivity degradation if it has failed due to low battery and hardware and software issues. In Fig. \ref{fig1}, UAV$_y$ is not reliable since it has many connections to the network, thereby could be critical and fail at any time. Therefore, the data of UE should not be solely transmitted  to UAV$_y$. Therefore, the RIS can concurrently boost the signal dedicated to the more reliable UAV$_x$ and relatively enhance the QoS of UAV$_y$. This process involves designing certain RIS elements’ phases to significantly maximize $\text{UE} \rightarrow \text{UAV}_{x}$ channel while a few elements’ phases should be configured to align with the $\text{UE} \rightarrow \text{UAV}_{y}$ channel. The two-UAV setup is considered for analytical tractability and to draw important insights into the RIS partitioning and deployment design.  

The number of RIS elements allocated for UAV$_j$ is denoted  by $N_j= \lceil\rho_jN\rceil$, where $\rho_j \in [0, 1]$ is the RIS allocation factor such that $\sum_{j=1}^2 \rho_j \leq 1$ and $\sum_{i=1}^2 N_i=N$ ensure that the total number of the allocated RIS elements for partitioning is equal to the physically available  number of the RIS elements. Furthermore, the reflection coefficient matrix of the RIS is modeled as $\mathbf{\Theta} =diag(G)$, where $G_n= |G_n|e^{j\theta_n}$ and $\theta_n \in [0, 2\pi)$ is the phase shift of the $n$-th element, with $|G_n|=1$, $\forall n$.

\subsection{Channel Model}
The channels between the UE and the UAVs (RIS), between the UAVs and the UAVs, and between the RIS and the UAVs  are considered to be quasi-static with flat-fading and presumed to be perfectly-known. To practically model the communication links between the UE and the RIS,  between the RIS and the UAVs, and between the UE and the UAVs, we consider the Nakagami$-m$ fading model that characterizes different fading conditions, i.e., ($m=1$) for severe fading and $(m=5)$ for nearly line-of-sight (LoS).
We represent the small-scale fading coefficient and path-loss for the $\text{UE} \rightarrow \text{RIS}$ channel as  $\mathbf g^\text{UR}$ and $\beta^\text{UR}$, respectively. Likewise, the small-scale fading coefficient and large-scale path-loss for the $\text{RIS} \rightarrow \text{UAV}_{j\in \{x,y\}}$ channel are denoted as $\mathbf g^\text{RK}_j$ and $\beta^\text{RK}_{j}$, respectively. Here, $\mathbf g^\text{UR}= [g^\text{UR}_{n}, \ldots, g^\text{UR}_{N}]$ and $\mathbf g^\text{RK}_{j}= [g^\text{RK}_{n,j}, \ldots, g^\text{RK}_{N,j}]$. Therefore, we consider the channel from the UE to UAV$_j$ over the $n$-th
RIS element as $g^\text{URK}_{n,j}= g^\text{UR}_{n} g^\text{RK}_{n,j}$, where $g^\text{UR}_{n}=|g^\text{UR}_{n}|e^{-j\phi_{n}}$ denotes the channel coefficient between the UE and the $n$-th RIS element, while  $g^\text{RK}_{n,j}=|g^\text{RK}_{n,j}|e^{-j\psi_{n,j}}$ is the channel coefficient
between the $n$-th RIS element and UAV$_j$; $|g^\text{UR}_{n}|$ and $|g^\text{RK}_{n,j}|$ are the channel amplitudes, while  $\phi_{n}$ and  $\psi_{n,j}$ are the channel phases.  

Moreover, for the direct links between the UE and the UAVs, let $g^\text{UK}_{k}$ and $\beta^\text{UK}_{k}$ denote the small-scale fading coefficient and path-loss for the $\text{UE} \rightarrow \text{UAV}_k$ channel, respectively. For the $\text{UE} \rightarrow \text{UAV}_k$ channel, the signal-to-noise ratio (SNR) is defined as  
$\gamma^\text{(UK)}_{k}=\frac{p|\sqrt{\beta^\text{UK}_{k}}g^\text{UK}_{k}|^2}{N_0}$, where $N_0$ is the additive white Gaussian noise (AWGN) variance. A typical UAV$_k$ is assumed to be within the transmission range of the UE if $\gamma^\text{(UK)}_{k} \geq \gamma^\text{UE}_{0}$, where $\gamma^\text{UE}_{0}$ is the minimum SNR threshold for the  UE-UAV communication links. 

For simplicity, we only consider the LoS path component for each UAV-UAV link, which is well justified for UAV communications  \cite{8292633, 9293155}.  Thus, the LoS path-loss between UAV$_k$ and UAV$_{k'}$ can be expressed as
 $\Gamma_{k,k'}=20\log\bigg( \frac{4 \pi f_c d_{k,k'}}{c} \bigg),$ where $d_{k,k'}$ is the distance between UAVs $k$ and $k'$, $f_c$ is the carrier frequency, $c$ is the speed of light, and $d_{k,k'}$ is the distance between UAV$_k$ and UAV$_{k'}$. The SNR in dB between UAV$_k$ and UAV$_{k'}$  is $\gamma^\text{(UAV)}_{k,k'}=10\log P-\Gamma_{k,k'}-10 \log N_0$.  UAV$_k$ and UAV$_{k'}$ have a successful connection provided that $\gamma^\text{(UAV)}_{k,k'} \geq \gamma^\text{UAV}_{0}$, where $\gamma^\text{UAV}_{0}$ is the minimum SNR threshold for the UAV-UAV communication links.

 \subsection{SNR Formulation}\label{SNR}
\begin{figure}[t!]  
\begin{center}
\includegraphics[width=0.95\linewidth, draft=false]{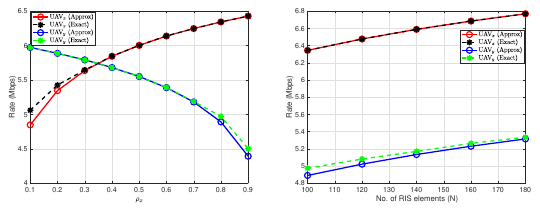}
\caption{Rate performance versus:  (left) $\rho_{x}$ and (right) number of RIS elements $N$ for $\rho_{x}=0.8$ and $b=4$.}
   \label{fig2}
\end{center}
 
\end{figure}

 In light of the preceding discussions, the SNR at UAV$_j$, $j \in \{x,y\}$,  can be expressed as
\begin{align}\label{exact}
 \gamma_{j}(\boldsymbol{\alpha})=\frac{p \beta^\text{UR}(\boldsymbol{\alpha})\beta^\text{RK}_{j}(\boldsymbol{\alpha})\bigg| \underbrace{\sum_{n=1}^{N_j}g^\text{URK}_{n,j}e^{j\theta_{n}}}_\textbf{aligned signal}+\underbrace{\sum_{\Tilde{n}=N_j+1}^{N}g^\text{URK}_{\Tilde{n},j}e^{j\theta_{\Tilde{n}}}}_\textbf{non-aligned signal}\bigg|^2}{N_0},   
\end{align}
where $\boldsymbol{\alpha}=[\alpha_x, \alpha_y, \alpha_z]$ is the Cartesian
coordinates of the RIS. Furthermore, due to the available perfect global
channel state information (CSI) and using high bit resolution of the RIS element’s phase shifter \cite{38}, we consider that a portion of RIS elements is
perfectly aligned with $\text{UE} \rightarrow \text{UAV}_{j=x}$ cascaded channel, while the rest RIS elements are aligned with $\text{UE} \rightarrow \text{UAV}_{j=y}$ channel
and not aligned with the $\text{UE} \rightarrow \text{UAV}_{j=x}$ channel. Hence, for the sake
of analytical tractability, we ignore the impact of non-aligned links from \eqref{exact}\footnote{Although we ignore the impact of non-aligned channels for tractability purposes, we show in Fig. \ref{fig2} that the impact of non-aligned channels on the SNR performance is negligible.}. 
 
 Moreover, for feasible optimization in this paper, we can represent the RIS elements’ allocation portions $\boldsymbol{\rho}=\{\rho_1, \rho_2\}$ to denote the RIS section
allocated to the UAVs$_{j}$, $j \in \{x,y\}$. Hence, $N_j$ in the first summation
term in \eqref{exact} can also be presented as $N_j=\lceil\rho_jN\rceil$. Additionally and without loss of generality, we rewrite the summation term of $\sum_{n=1}^{\lceil\rho_jN\rceil}(.)$ as $\rho_j \sum_{n=1}^{N}(.)$\footnote{Note that real-valued RIS portions may result in non-integer RIS partition allocation. In this case, the RIS   allocates some elements  for exact portion allocated for UAVs $j \in \{x,y\}$.}.  We further
rewrite  $ \gamma_{j}(\boldsymbol{\alpha}, \boldsymbol{\rho})$ as
\begin{align}\label{appro}
  &\gamma_{j}(\boldsymbol{\alpha}, \boldsymbol{\rho})= \frac{p \beta^\text{UR}(\boldsymbol{\alpha})\beta^\text{RK}_{j}(\boldsymbol{\alpha})\bigg|\rho_j \sum_{n=1}^{N}g^\text{URK}_{n,j}e^{j\theta_{n}}\bigg|^2}{N_0} \nonumber  \\ & \nonumber =\frac{p \beta^\text{UR}(\boldsymbol{\alpha})\beta^\text{RK}_{j}(\boldsymbol{\alpha})\bigg|\rho_j \sum_{n=1}^{N}|g^\text{UR}_n||g^\text{RK}_{n,j}|e^{j(\theta_{n}-\phi_n-\psi_{n,j})}\bigg|^2}{N_0} \\&  =\frac{p \beta^\text{UR}(\boldsymbol{\alpha})\beta^\text{RK}_{j}(\boldsymbol{\alpha})\rho_j^2 |Q|^2}{N_0},
\end{align}
where $Q=\sum_{n=1}^N Q_n=\sum_{n=1}^N|g^\text{UR}_n||g^\text{RK}_{n,j}|$ is the $N$-element double-Nakagami$-m$ that is independent and
identically distributed (i.i.d.) random variable (RV) with parameters $m_1$, $m_2$, $\Omega_1$, and $\Omega_2$, i.e., the distribution of the product
of two RVs following the Nakagami$-m$ distribution with the probability density function (PDF) is given in \cite{PLS}.

The total channel gain for $\text{UE} \rightarrow \text{UAV}_{j}$ cascaded channels in \eqref{appro} can be maximized through the optimization of the reflection matrix $\mathbf{\Theta}$. Since RIS's controller has perfect knowledge of the CSI, it has the capability to determine suitable phase shifts that can effectively nullify the phases of the respective $\text{UE} \rightarrow \text{RIS}$ and $\text{RIS} \rightarrow \text{UAV}_{j}$ channels as $\theta_n=\phi_n+\psi_{n,j}$, $\forall n, j \in \{x,y\}$ \cite{PLS, 38}. Specifically, for the $n$-th RIS element, given $\phi_n$ and $\psi_{n,j}$, the RIS's controller can perfectly align $\theta_n$ with $\phi_n$ and $\psi_{n,j}$ to nullify their effect, which provides the maximum channel gain value of $\text{UE} \rightarrow \text{UAV}_j$ link.

In the remaining of this subsection, we present some numerical results to assess the
equivalence of the analytical SNR expressions in \eqref{exact} and \eqref{appro}. All channels are modeled with Nakagami$-m$ fading distribution. Therefore, we set $m=1$ to emulate the Rayleigh channel for the  $\text{UE} \rightarrow \text{UAV}_{k}$ links, and $m_1=5$, and $m_2=1$, $\Omega_1=\Omega_2=1$ for $\text{UE} \rightarrow \text{RIS}$ and $\text{RIS} \rightarrow \text{UAV}_j$ links, respectively. Unless stated otherwise, the following system parameters are considered: $p=23$ dBm, $P=30$ dBm, $N=100$, $N_0=-120$ dBm. For the purposes of
this part, the fixed 3D Cartesian coordinates in meters for the UE are $(318, 220, 0)$, while  UAV$_x$, UAV$_y$, and the RIS are located at $(460, 340, 200)$, $(370, 14, 200)$, and $(0, 0, 120)$, respectively.

Fig. \ref{fig2} compares exact and approximated rates
averaged over $10^5$ Matlab simulations, which are calculated using the SNRs in \eqref{exact} and \eqref{appro} with bandwidth $B=250$ KHz. For UAV$_j$, the RIS partitions, respectively, are $\rho_{x}$ and  $\rho_{y}=1-\rho_{x}$.  We consider phase shift control using phase shift quantization levels similar to \cite{38}, where parameter $b$ is the bit resolution of the RIS element’s phase shifter. In the literature \cite{38}, $b=4=\infty$ is considered for high bit resolution (perfect phase shift). From the figure, we notice that approximated rates for the UAVs tightly match the exact rates for most values of $\rho_{x}$ and $N$. Overall, these results justify our assumption to ignore the impact of non-aligned channels from the other portions of the RIS. Therefore, in the remaining of this paper, we will use the approximated SNR expression in \eqref{appro}. 


\section{Problem Formulation}

\subsection{Network Connectivity}
We model the uplink RIS-assisted UAV model using the graph network $\mathcal G(\mathcal V, \mathcal E)$, where $\mathcal V$ represents the set of vertices associated with the network nodes (UAVs and UE) and $\mathcal E$ represents the edges (links). For a graph edge $e_{l}$, $ 1 \leq l \leq E$, that links two nodes $(u, v) \in \mathcal V$, we have:
\begin{equation}
e_l = \begin{cases}
1 & \text{if}  ~  \gamma^\text{(UAV)}_{u,v} \geq \gamma_0^\text{UAV} ~ \text{for}~ \text{UAV}_u \rightarrow \text{UAV}_{v}~ \text{link},\\
1 & \text{if}  ~  \gamma^\text{(UK)}_{v} \geq \gamma_0^\text{UE} ~ \text{for}~ \text{UE} \rightarrow \text{UAV}_{v} ~ \text{link}, \\
0 & \text{otherwise}.
\end{cases}
\end{equation}
Subsequently, the weight vector $\mathbf w \in {[\mathbb R^+]}^E$ of these links is defined as $\mathbf w= [ w_1, w_2, \ldots, w_E]$, and is given element-wise as
\begin{equation}
w_l = \begin{cases}
\gamma^\text{(UAV)}_{u,v} & ~\text{for}  ~  \text{UAV}_u \rightarrow \text{UAV}_{v}~ \text{link},\\
\gamma^\text{(UK)}_{v} &   ~ \text{for}~ \text{UE} \rightarrow \text{UAV}_{v} ~ \text{link}.
\end{cases}
\end{equation}
For $e_{l}$, let $\mathbf a_l$ be a vector, where the $u$-th and $v$-th elements in $\mathbf a_l$ are given by $a_{u,l}=1$ and $a_{v,l}=-1$, respectively, and zero otherwise. Let $\mathbf A$ be the incidence matrix of a graph $\mathcal G$ with the $l$-th column given by $\mathbf a_l$.  Hence, in an undirected graph $\mathcal G(\mathcal V, \mathcal E)$, the Laplacian matrix $\mathbf M$ is a $V \times V$ matrix, defined as \cite{4657335}:
\begin{equation} \label{lap}
\mathbf M= \mathbf A  ~diag(\mathbf w) ~\mathbf A^T=\sum^{E}_{l=1} w_l \mathbf a_l \mathbf a^T_l,
\end{equation}
where the entries of $\mathbf M$ are given element-wise by
\begin{equation}
M(u,v) = \begin{cases}
deg(u) &\text{if} ~u=v,\\
-w_{l} &\text{if}~ (u, v) \in \mathcal E, \\
0 & \text{otherwise},
\end{cases}
\end{equation}
where $deg(u)$ is the degree of node $u$, which represents the number of its neighboring nodes. 

To maximize the connectivity of the uplink RIS-assisted UAV network, we choose a well-known metric, known as the \textit{algebraic connectivity} \cite{new, 4657335, 9593204, 4786516, 8292633, saifglobecom}, denoted as $\lambda_2(\mathbf M)$, to measure how well a network is connected.  With RIS deployment and partitioning, a new graph $\mathcal G'(\mathcal V, \mathcal E')$ is constructed with the same number of $V$ nodes and a larger set of edges denoted by $\mathcal E'$ with $\mathcal E'=\mathcal E \cup \mathcal E_{new}$, where $\mathcal E_{new}$ is the new edges for the  
$\text{UE} \rightarrow  \text{UAV}_{x}$ and $\text{UE} \rightarrow  \text{UAV}_{y}$ links. The gain can be realized by computing $\lambda_2 (\mathbf M') \geq \lambda_2 (\mathbf M)$, where $\mathbf M'$ is the resulting Laplacian matrix of the new graph $\mathcal G'(\mathcal V, \mathcal E')$.

In this paper, we measure the reliability of the UAVs$_{j\in \{x,y\}}$ based on the severity of network connectivity after removing UAV$_j$ and its connected edges to other nodes, which is defined  as $\mathcal R_j=\frac{1}{\lambda_2(\mathcal G_{-j})}$, where $\mathcal G_{-j}$ is the sub-graph resulting from removing UAV$_j$ and all its adjacent edges to other nodes in $\mathcal G$. Thus, we consider $\mathcal R_y>\mathcal R_x$.

\subsection{Problem Formulation}
In this paper, our interest is to unleash the benefits of the RIS  to aid the uplink connections of the UE to the blocked UAVs. Following the previous results and discussions suggest, the SNR performance improvement inherently requires the joint optimization of RIS partitioning and location to maximize the link quality between the UE and the UAVs via the RIS while ensuring their QoS constraints. Let  $\gamma^\text{RIS}_{0}$ be the minimum SNR threshold of  $\text{UE} \rightarrow \text{UAV}_{j\in\{x,y\}}$  via the RIS. The considered optimization problem is formulated as  
\begin{subequations} \nonumber 
\label{eq10}
\begin{align}
&\textbf{P0:} ~\max_{\boldsymbol{\rho}, \boldsymbol{\alpha}} ~~~~ \lambda_2(\mathbf M' (\boldsymbol{\rho}, \boldsymbol{\alpha}))
\label{eq10a}\\
 &~~~~~~~~~~{\rm s.~t.\ }\\
 &\text{$C_1^0$:} ~~~~~~~~~~\gamma_x(\boldsymbol{\rho}, \boldsymbol{\alpha})\geq \gamma^\text{RIS}_{0},\\
 & \text{$C_2^0$:} ~~~~~~~~~~\gamma_y(\boldsymbol{\rho}, \boldsymbol{\alpha})\leq \zeta\gamma^\text{RIS}_{0},\\
& \text{$C_3^0$:}~~~~~~~~~~\sum_{j=1}^2\rho_j\leq 1, \tag{7}\\
&\text{$C_4^0$}:~~~~~~~~~~  0 \preceq \boldsymbol{\rho} \preceq 1, -\infty \preceq \boldsymbol{\alpha} \preceq \infty,
\end{align}
\end{subequations} 
where $\preceq$ is the pairwise inequality and $0 < \zeta \leq 1$ is a design parameter that determines the QoS limit set for the least reliable UAV$_y$ based on $\mathcal R_y$.
In \textbf{P0}, \text{$C_1^0$} and \text{$C_2^0$}
constitute
the QoS constraints on the UAVs$_{j \in \{x,y\}}$,  \text{$C_3^0$} ensures that the allocated portions does not exceed unity to limit the total number of allocated RIS elements is not higher than the total number of RIS elements. Finally, \text{$C_4^0$} specifies the domain of optimization variables. The Laplacian matrix $\mathbf M' (\boldsymbol{\rho}, \boldsymbol{\alpha})$ depends on the RIS location and partitions, which determine the quality of the new links. To tackle \textbf{P0} over $\boldsymbol{\rho}$ and $\boldsymbol{\alpha}$, we decompose \textbf{P0} into two subproblems and solve them  iteratively. For given RIS location $\boldsymbol{\alpha}_0$, we optimize $\lambda_2(\mathbf M' (\boldsymbol{\rho}, \boldsymbol{\alpha}_0))$ by finding a closed-form solution of the RIS portions, denoted by $\boldsymbol{\rho}^*$,  that pushes the  UAVs$_{j \in \{x,y\}}$ SNR to its maximum. Then, we exploit the closed-form RIS partitioning solution to find the 3D deployment of the RIS to further maximize  $\lambda_2(\mathbf M' (\boldsymbol{\rho}^*, \boldsymbol{\alpha}))$.

\section{Proposed Solution}
Maximizing network connectivity requires optimization over a graph structure, i.e., the selection of UE and UAV nodes  and their link quality. Since the nodes (the UE and the UAVs$_{j \in \{x,y\}}$) are known and network connectivity is a monotonically
increasing function of the added links and their weights \cite{new}, we can equivalently maximize
the network connectivity by maximizing the SNR of the added new $\text{UE} \rightarrow \text{UAV}_{j\in \{x,y\}}$ links via the RIS.

\subsection{RIS Partitioning Optimization}
The first subproblem of the RIS partitioning to
maximize the sum SNR of the UAVs$_{j \in \{x,y\}}$ can be
formulated for a given RIS location $\boldsymbol{\alpha}_0$ by using \textbf{P0} and \eqref{appro}  as follows 
\begin{subequations} \nonumber 
\label{eq10}
\begin{align}
&\textbf{P1:} ~\max_{0 \preceq \boldsymbol{\rho} \preceq 1} ~~~~  \gamma_x(\boldsymbol{\rho}, \boldsymbol{\alpha}_0)+\gamma_y(\boldsymbol{\rho}, \boldsymbol{\alpha}_0)
\label{eq10a}\\
 &~~~~~~~~~~{\rm s.~t.\ }\\
 &\text{$C_1^1$:} ~~~~~~~~~~~~~~\gamma_x(\boldsymbol{\rho}, \boldsymbol{\alpha_0})\geq \gamma^\text{RIS}_{0},\\
& \text{$C_2^1$:}~~~~~~~~~~~~~~\gamma_y(\boldsymbol{\rho}, \boldsymbol{\alpha_0})\leq \zeta\gamma^\text{RIS}_{0}, \tag{8}\\
& \text{$C_3^1$:}~~~~~~~~~~~~~~\sum_{j=1}^2\rho_j\leq 1,
\end{align}
\end{subequations} 
which can be solved by pushing the SNR of the $\text{UE} \rightarrow  \text{UAV}_{x}$ link, corresponding to the most reliable UAV, to its maximum value.
In what follows, we provide the closed-form solution for the RIS partitioning.

First, the sum SNR can be maximized by
satisfying UAV $y$'s QoS constraint in \text{$C_2^1$}
with equality while
pushing $\gamma_x$ to its maximum subject in \text{$C_1^1$}. Therefore, we
rewrite \text{$C_2^1$} 
in (8) using \eqref{appro} as follows
\begin{align}\label{closed}
\gamma_{y}(\boldsymbol{\rho}, \boldsymbol{\alpha_0})&=\frac{p \beta^\text{UR}(\boldsymbol{\alpha_0})\beta^\text{RK}_{y}(\boldsymbol{\alpha_0})\rho_{y}^2 |Q|^2}{N_0}= \zeta \gamma^\text{RIS}_0.
\end{align}
The closed-form solution
for $\rho_{y}$ can be derived as  
\begin{align}\label{closedy}
 \rho^*_{y}(\boldsymbol{\alpha_0})=\sqrt{\frac{\zeta \gamma^\text{RIS}_0 N_0}{p \beta^\text{UR}(\boldsymbol{\alpha_0})\beta^\text{RK}_{y}(\boldsymbol{\alpha_0})|Q|^2}}. 
\end{align}
Subsequently, the rest of the unused elements should be allocated to
UAV$_{x}$, and therefore, all available RIS elements must be used to reach the
maximum SNR of that UAV. Thus, the RIS portion allocated to UAV$_{x}$ is
written as
\begin{align}\label{closedx}
 \rho^*_{x}(\boldsymbol{\alpha_0})=1-  \rho^*_{y}(\boldsymbol{\alpha_0}).
\end{align}
Notice that the solution $ \rho^*_{y}(\boldsymbol{\alpha_0})$ and $ \rho^*_{x}(\boldsymbol{\alpha_0})$  are feasible only if the constraints in (8) are satisfied. In general, the RIS partitioning closed-form solution of the two-UAV setup can be generalized to multiple-UAV setup, where  $\rho^*_{x}$ is given as 
\begin{align}\label{closedx}
 \rho^*_{x}(\boldsymbol{\alpha_0})=1-  \sum_{i=1}^{C}\rho^*_{i}(\boldsymbol{\alpha_0}),
\end{align}
where $C$ is any number of blocked UAVs that is greater than one. The solution first needs to find the RIS partitions that satisfy the SNR limit of the least reliable UAV and so on.

\subsection{RIS Deployment}
The RIS location substantially affects the network connectivity performance as it directly relates to the channel gains and path-loss for the $\text{UE} \rightarrow \text{UAV}_{j\in \{x,y\}}$ links via the RIS. Therefore, this subsection considers the RIS deployment problem formulated in \textbf {P2} to maximize the SNR of those links  by using the RIS portions derived in the previous subsection:
\begin{subequations} \nonumber 
\label{eq10}
\begin{align}
&\textbf{P2:} ~\max_{\boldsymbol{\alpha}} ~~~~  \gamma_x(\boldsymbol{\rho^*}, \boldsymbol{\alpha})+\gamma_y(\boldsymbol{\rho^*}, \boldsymbol{\alpha})
\label{eq10a}\\
 &~~~~~~~~~~{\rm s.~t.\ }\\
 &\text{$C_1^2$:} ~~~~~~~~~~\gamma_x(\boldsymbol{\rho^*}, \boldsymbol{\alpha})\geq \gamma^\text{RIS}_{0},\\
& \text{$C_2^2$:}~~~~~~~~~~\gamma_y(\boldsymbol{\rho^*}, \boldsymbol{\alpha})\leq \zeta\gamma^\text{RIS}_{0},\\
&\text{$C_3^2$:}~~~~~~~~~~  -\infty \preceq \{\alpha_x, \alpha_y\}\preceq \infty,  0 \preceq \{\alpha_z\}\preceq \infty,
\end{align}
\end{subequations} 
where $\boldsymbol{\rho^*}=[\rho^*_{j=x}, \rho^*_{j=y}]$. \text{$C_3^2$}
defines the search space range for the  $\alpha_x, \alpha_y$
coordinates as extending from negative infinity to positive infinity, while the coordinate constraint for $\alpha_z$ is limited to a range between $0$ and $\infty$. We solve \textbf{P2} by meta-heuristic
methods that run a global search of the RIS locations and evaluate the location fitness by the proposed RIS partitioning. In this work, we use the simulated-annealing (SA) method to solve \textbf{P2}. 

 \section{Numerical Results}\label{NR}
In this section, we present the numerical results for solving $\mathbf P_0$, where system parameters are the same as in \sref{SNR} for $2$ UAVs in the network. Unless otherwise stated, $N=100$, $K=8$, and $\zeta=0.2$, which means that the QoS for $\text{UE} \rightarrow \text{UAV}_{y}$ link is enforced to be $2$\% of the QoS for $\text{UE} \rightarrow \text{UAV}_{x}$ link. In these
simulations, we utilize the 3GPP Urban Micro (UMi) model \cite{PLS} at a carrier frequency of $3$ GHz to compute the large-scale path loss values for UE-UAVs links, while similar to \cite{saifglobecom}, we use $\sqrt{\frac{\beta_0}{(d^\text{UR})^2}}$ and $\sqrt{\frac{\beta_0}{(d_{j}^\text{RK})^2}}$ for $\text{UE} \rightarrow \text{RIS}$ and $\text{RIS} \rightarrow \text{UAV}_{j}$ links, respectively, where $\beta_0$ denotes the path loss at the reference distance $d_\text{ref}=1$ m  and $d$ is the corresponding distance.

\begin{figure}[t!]  
\begin{center}
\includegraphics[width=0.75\linewidth, draft=false]{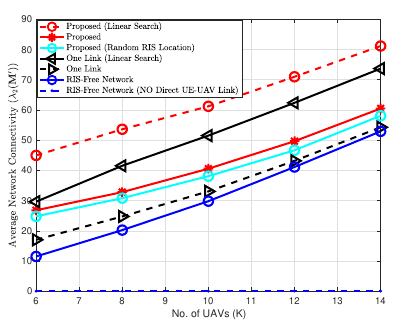}
\caption{Network connectivity versus the number of UAVs $K$ for $N=100$, $\gamma^\text{RIS}_{0}=60$ dB, and $\zeta=0.2$.}
   \label{fig3}
\end{center}
 
\end{figure}

Fig. \ref{fig3} compares the network connectivity of the proposed scheme versus the number of UAVs $K$  with the following benchmark scenarios. 1) One Link: All RIS elements are allocated to beamform the UE's signal to a predefined UAV$_x$; 2) One Link (Optimal): All RIS elements are allocated to beamform the UE's signal to the selected optimal UAV, which is found  via linear search; 3) Proposed (Random RIS Location): It is the  proposed scheme but with a fixed RIS
 location; 4) RIS-Free Network; 5) RIS-Free Network (No Direct UE-UAV Link). Notice that the one link schemes are special cases of the proposed (optimal) and proposed schemes, where RIS elements are not allocated to align with UAV$_y$, thus the QoS constraint in \text{$C_2^1$}
is satisfied. The results show that the proposed (optimal) scenario
leads to significant connectivity performance  with $80$ at $K=14$. This is due to the fact that the proposed (optimal) scenario exploites the full benefit of RIS partitioning and deployment to create two links while finding the desired two blocked UAVs that maximizes the connectivity; finding the desired two blocked UAVs is via linear search of all the $2$ UAV combinations. For this reason, the proposed (optimal) is better than the proposed, where we predefined UAV$_x$ and UAV$_y$ in advance.
On the other hand, RIS-free network scenario provides the worst connectivity. Furthermore, we can notice that
the  proposed (random RIS location) scenario slightly decreases the connectivity. For example, when $K$ is $10$, the connectivity of proposed (random RIS location) is $38$, while that for  proposed is $40$. Finally, the RIS-free network (no direct UE-UAV link) scheme has zero connectivity since the network graph is not one component. This showcases the leverage of deploying the RIS for maximizing connectivity of UAV networks not to only support UE communication, but also to connect the UE to the network when the direct UE-UAV link is unavailable.

\begin{figure}[t!]  
\begin{center}
\includegraphics[width=0.75\linewidth, draft=false]{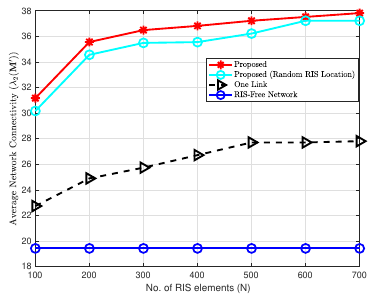}
\caption{Network connectivity versus the number of RIS elements $N$ for $K=8$, $\gamma^\text{RIS}_{0}=60$ dB, and $\zeta=0.2$.}
   \label{fig4}
\end{center}
 
\end{figure}

Fig. \ref{fig4} compares the network connectivity versus RIS elements $N$. It is
noticed that the increase in the number of RIS elements increases connectivity, since  RIS with more elements can boost up the quality of the new $\text{UE} \rightarrow \text{UAV}_{x}$ and $\text{UE} \rightarrow \text{UAV}_{y}$ links. Specifically, since our RIS partitioning closed-form solution calculates $\rho_{y}$ that satisfies the minimum QoS of $\text{UE} \rightarrow \text{UAV}_{y}$ link, the remaining RIS elements, which increases as $N$ increases, are configured for  $\text{UE} \rightarrow \text{UAV}_{x}$ link to maximize its SNR to the maximum value. Thus, this increases the network connectivity. RIS-free scheme is maintained fixed since it has nothing to do with changing the RIS elements.

\begin{figure}[t!]  
\begin{center}
\includegraphics[width=0.95\linewidth, draft=false]{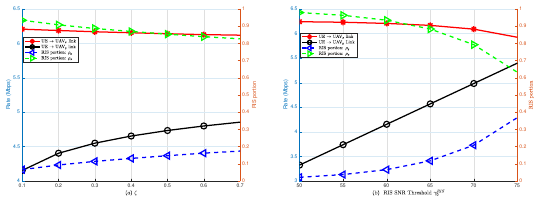}
\caption{Rates of $\text{UE} \rightarrow \text{UAV}_{x}$ and $\text{UE} \rightarrow \text{UAV}_{y}$ links and the corresponding RIS portions versus (a) $\zeta$ for $\gamma^\text{RIS}_{0}=60$ dB and (b) $\gamma^\text{RIS}_{0}$ for $\zeta=0.1$.}
   \label{fig5}
\end{center}
 
\end{figure}

Fig. \ref{fig5}(a) demonstrates the rate performance, which is calculated as $B\log_2(1+\gamma_j(\boldsymbol{\alpha},\boldsymbol{\rho}))$, versus $\zeta$ for $N=100$ and $\gamma^\text{RIS}_{0}=60$ dB. It is noticed that the increase in $\zeta$ leads to increase the QoS threshold of the $\text{UE} \rightarrow \text{UAV}_{y}$ link, which needs more RIS elements to satisfy it,  thus decreases the SNR for $\text{UE} \rightarrow \text{UAV}_{x}$ link, as expected. For example, when $\zeta=0.3$, RIS portion for $y$ case is about $0.1$ to satisfy SNR of $0.3*\gamma^\text{RIS}_{0}$, while RIS portion for $x$ case is about $0.9$ to satisfy the maximum SNR of the $\text{UE} \rightarrow \text{UAV}_{x}$ link. Accordingly, the rate of the $\text{UE} \rightarrow \text{UAV}_{y}$ link is $4.4$ Mbps, while for the $\text{UE} \rightarrow \text{UAV}_{x}$ link is $6.3$ Mbps. This decrease in UAV$_x$ rate is
because more RIS elements are required to satisfy the increasing QoS requirement for the  $\text{UE} \rightarrow \text{UAV}_{y}$ link. As a result, fewer RIS elements are left to boost up the signal of $\text{UE} \rightarrow \text{UAV}_{x}$ link.

Fig. \ref{fig5}(b) plots the rate performance  versus $\gamma^\text{RIS}_{0}$ for $\zeta=0.1$ and $N=100$. In terms of RIS allocations, we observe that the increase of the added links QoS via increasing $\gamma^\text{RIS}_{0}$ leads to significant increase  in  the number of RIS elements needed to satisfy the minimum QoS requirement for  $\text{UE} \rightarrow \text{UAV}_{y}$ link. For example, when $\gamma^\text{RIS}_{0}$ increases from $60$ dB to $65$ dB, $\rho_{y}$ jumps from  $0.0655$ (around $7$ elements) to  $0.1167$ ($11$ elements), and almost $30\%$ of the RIS elements is partitioned to align with UAV$_y$ when $\gamma^\text{RIS}_{0}=75$ dB. Thus, the rate of  $\text{UE} \rightarrow \text{UAV}_{y}$ link increases and the rate of  $\text{UE} \rightarrow \text{UAV}_{x}$ link decreases since a few RIS elements are left to be aligned with UAV$_x$.

\section{conclusion}
In this paper, we have studied the UAV connectivity by exploiting RIS partitioning and deployment to improve connectivity of UAV networks,  supporting UE-UAV communication or connecting the UE to the network when direct links are unavailable. We have developed a closed-form solution for RIS partitioning, while simulated-annealing is used  for RIS deployment. Simulation results have shown that with the introduction of RIS partitioning, substantially higher achievable connectivity and SNR  can be obtained compared to its RIS-free  and one link counterpart.

\end{document}